\documentclass[twocolumn,english,showpacs,preprintnumbers,prbr]{revtex4-1}
\usepackage[latin9]{inputenc}
\usepackage{amsmath}
\usepackage{amssymb}
\usepackage{graphicx}
\usepackage{esint}
\usepackage{natbib}
\usepackage{hyperref}
\hypersetup{colorlinks=true,linkcolor=blue,filecolor=magenta,urlcolor=blue,citecolor=blue,}

\makeatletter

\@ifundefined{textcolor}{}{%
 \definecolor{BLACK}{gray}{0}
 \definecolor{WHITE}{gray}{1}
 \definecolor{RED}{rgb}{1,0,0}
 \definecolor{GREEN}{rgb}{0,1,0}
 \definecolor{BLUE}{rgb}{0,0,1}
 \definecolor{CYAN}{cmyk}{1,0,0,0}
 \definecolor{MAGENTA}{cmyk}{0,1,0,0}
 \definecolor{YELLOW}{cmyk}{0,0,1,0}
}

\usepackage{color}
\usepackage{ulem}

\usepackage{aecompl}

\usepackage{epsfig}\usepackage{dcolumn}\usepackage{bm}

\usepackage{babel}

\makeatother

\usepackage{babel}
\begin{document}

\title{Antibonding Ground state of Adatom Molecules in Bulk Dirac Semimetals}

\author{Y. Marques$^{1}$, A. E. Obispo$^{2,3}$, L. S. Ricco$^{1}$, M. de Souza$^{2}$, I. A. Shelykh$^{4,5}$, and
A. C. Seridonio$^{1,2}$}

\affiliation{$^{1}$Departamento de F\'{i}sica e Qu\'{i}mica, Unesp - Univ Estadual
Paulista, 15385-000, Ilha Solteira, SP, Brazil\\
 $^{2}$IGCE, Unesp - Univ Estadual Paulista, Departamento de F\'{i}sica,
13506-900, Rio Claro, SP, Brazil\\
 $^{3}$Departamento de F\'{i}sica - Universidade Federal do Maranhão, 65080-805, São Luís, MA, Brazil\\
 $^{4}$Science Institute, University of Iceland, Dunhagi-3, IS-107,
Reykjavik, Iceland\\
 $^{5}$ITMO University, St. Petersburg 197101, Russia}
\begin{abstract}
The ground state of the diatomic molecules in nature is inevitably bonding and its first excited state is antibonding. We demonstrate theoretically that for a pair of distant adatoms placed buried in 3D-Dirac semimetals, this natural order of the states can be reversed and antibonding ground state occurs at the lowest energy of the so-called bound states in the continuum. We propose experimental protocol with use of STM-tip to visualize the topographic map of the local density of states on the surface of the system to reveal the emerging Physics.
\end{abstract}
\maketitle

\textit{Introduction.}---Three-dimensional Dirac semimetals (3D-DSMs) such as $\text{Cd}_{3}\text{As}_{2}$ and $\text{Na}_{3}\text{Bi}$\cite{DSM1,DSM2,DSM3,DSM4,DSM5} represent novel class of functional materials constituting 3D analogous of gapless graphene\cite{Graphe1,Graphe2,Graphe3}. The band structure of 3D semimetals contains the set of the so-called Dirac points in which conduction and valence bands touch and effective mass becomes zero. Around these points the dispersion of quasiparticles corresponds to those of massless relativistic Dirac particles which result in series of unusual properties of these materials such as linear magnetoresistance, unprecedented Shubnikov-de Haas oscillations and ultrahigh carrier mobility\cite{Feature1,Feature2,Feature3}.

\begin{figure}[!]
\includegraphics[scale=0.5]{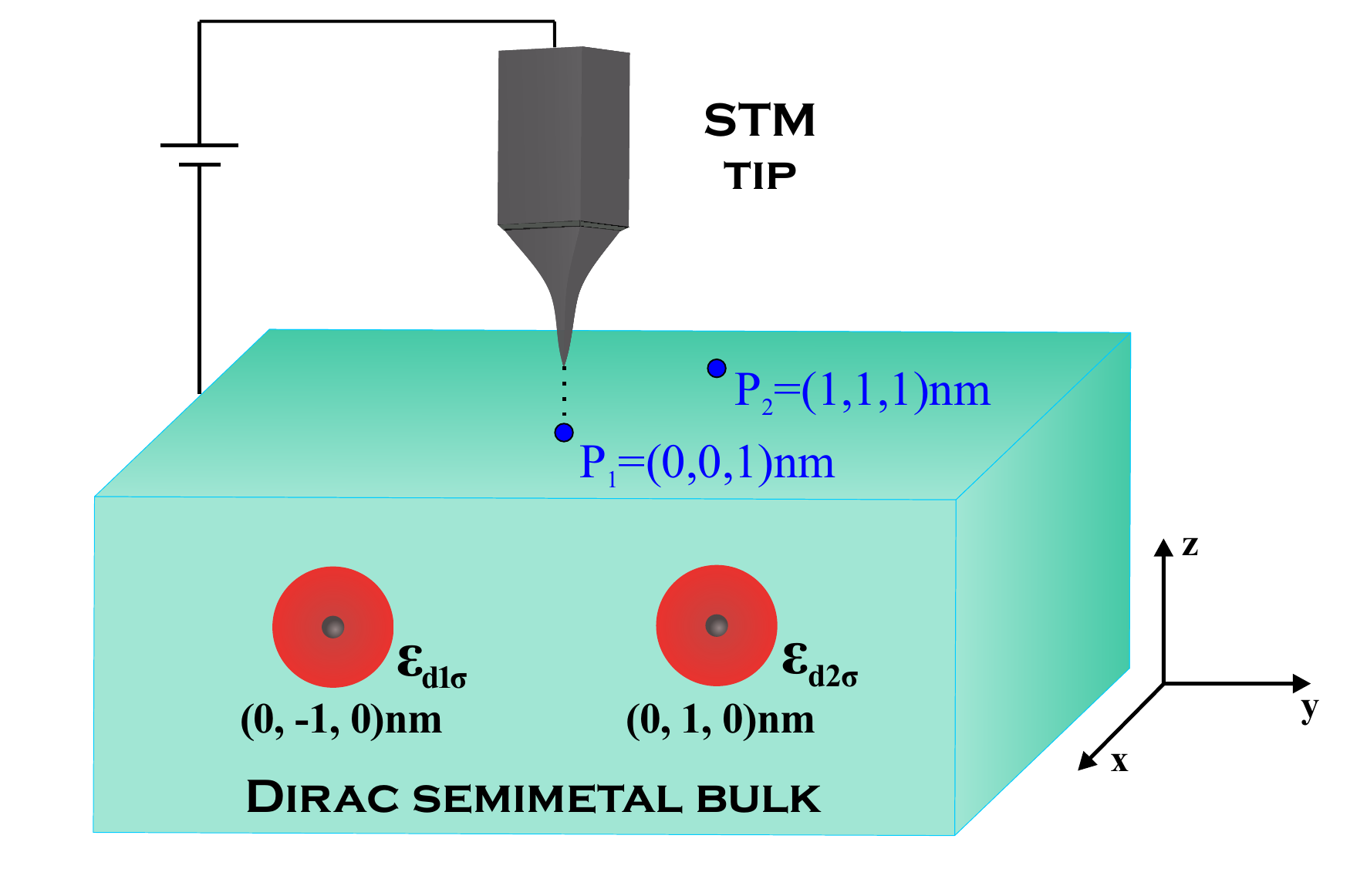} \protect\protect\protect\caption{\label{fig:Pic1}(Color online) Sketch of the setup proposed: two
adatoms with energy levels $\text{{\ensuremath{\varepsilon_{d_{1\sigma}}}}}$
and $\text{{\ensuremath{\varepsilon_{d_{2\sigma}}}}}$, buried in a 3D-DSM, at the positions $\mathbf{R}_{1}=(0,-1,0)\text{{nm}}$
and $\mathbf{R}_{2}=(0,1,0)\text{{nm}}$, respectively. P1 and P2 on top of the host represent sites in which the LDOS are probed by an STM-tip.}
\end{figure}

In this work, we predict one more interesting feature of such materials. Namely, if we consider buried pair of distant adatoms in the bulk of a 3D-DSM as depicted at Fig.\ref{fig:Pic1}, the ground state of this molecular system formed from {bound states in the continuum (BICs)} of the adatoms\cite{BIC1,BIC2,BIC3} will be characterized by antibonding-type orbital. This differs from the natural order in diatomic molecules where the ground state is of bonding-type in vast majority of cases {and formation of antibonding ground state till now was reported only in systems of artificially fabricated $\text{In}\text{As}$ and $\text{Ge}/\text{Si}$ \textit{p}-type quantum dots for certain values of the inter-dot separations\cite{Experiment,Experiment2}.} The behavior we report is a unique effect arising from long-range correlations between distant adatoms mediated by bulk fermions in 3D-DSMs. To detect the predicted effect, we propose to use the experimental approach developed in Ref.\cite{STM} for imaging isodensity contours of molecular states by scanning tunneling microscope (STM)-tip, as outlined at Fig.\ref{fig:Pic1}.

\textit{The Model.}---For theoretical analysis of two adatoms buried inside 3D-DSM, as depicted at Fig.\ref{fig:Pic1}, we employ an Anderson-like Hamiltonian\cite{Hamiltonian1,Hamiltonian2,Hamiltonian3}
\begin{equation}
\mathcal{H}_{\text{{T}}}=\mathcal{H}_{\text{{0}}}+\mathcal{H}_{\text{{d}}}+\mathcal{H}_{V},
\end{equation}
in which the effective low energy term describing the 3D-DSM is given by
\begin{eqnarray}
\mathcal{H}_{\text{{0}}} & = & \sum_{\mathbf{k},\tau}\psi_{\tau}^{\dagger}(\mathbf{k})h_{\tau}(\mathbf{k})\psi_{\tau}(\mathbf{k}),\label{eq:band}
\end{eqnarray}
where $\psi_{\tau}^{\dagger}(\mathbf{k})=(\begin{array}{cc}
c_{\mathbf{k}\tau\uparrow}^{\dagger} & c_{\mathbf{k}\tau\downarrow}^{\dagger})\end{array}$ is a spinor with fermionic operators $c_{\mathbf{k}\tau\sigma}^{\dagger}$
($c_{\mathbf{k}\tau\sigma}$) for creation (annihilation) of electrons
in quantum states labeled by the wave vector $\mathbf{k}$, spin $\sigma$ and chirality $\tau=\pm,$ and
\begin{eqnarray}
h_{\tau}(\mathbf{k}) & = & v_{F}\tau(k_{x}\sigma_{x}+k_{y}\sigma_{y}+k_{z}\sigma_{z}),\label{eq:ht}
\end{eqnarray}
where $\sigma_{i}$ accounts for the Pauli matrices and $v_{F}$ is the Fermi velocity.

The term
\begin{equation}
\mathcal{H}_{\text{{d}}}= \sum_{j\sigma}\varepsilon_{d_{j\sigma}}d_{j\sigma}^{\dagger}d_{j\sigma}+\sum_{j}U_{j}n_{d_{j\uparrow}}n_{d_{j\downarrow}}
\end{equation}
describes the buried adatoms ($j=1,2$), where $n_{d_{j\sigma}}=d_{j\sigma}^{\dagger}d_{j\sigma}$,
$d_{j\sigma}^{\dagger}$ ($d_{j\sigma}$) creates (annihilates) an electron with spin $\sigma$ in the state $\varepsilon_{d_{j\sigma}},$
and $U_{j}$ is the on-site Coulomb repulsion.

$\mathcal{H}_{V}$ accounts for the hybridization between the adatoms and the host,
\begin{equation}
\mathcal{H}_{V}=\sum_{j\mathbf{k}\tau}\hat{d}_{j}^{\dagger}\hat{V}_{j\mathbf{k}}\psi_{\tau}(\mathbf{k})+\text{H.c.},
\end{equation}
where $\hat{d}_{j}^{\dagger}=(\begin{array}{cc}
d_{j\uparrow}^{\dagger} & d_{j\downarrow}^{\dagger}\end{array})$ and
\begin{eqnarray}
\hat{V}_{j\mathbf{k}} & = & \left(\begin{array}{cc}
V_{j\mathbf{k}} & 0\\
0 & V_{j\mathbf{k}}
\end{array}\right)\label{eq:Vkmatrix}
\end{eqnarray}
is hybridization matrix. We assume that both adatoms are equally coupled to the 3D-DSM in such a way that
$V_{j\mathbf{k}}=\frac{v_{0}}{\sqrt{\mathcal{N}}}e^{i\mathbf{k}\cdot\mathbf{R}_{j}}$, in which $\mathcal{N}$ gives the total number of states in the band-structure and $\mathbf{R}_{j}$ corresponds to the positions of the buried adatoms.

To explore the effects induced by the adatoms, we focus on the local density of states of the system (LDOS) given by
\begin{equation}
\text{{LDOS}}(\varepsilon,{\bold R_{m}})=-\frac{1}{\pi}{\tt Im}[\sum_{\sigma}\tilde{\mathcal{G}}_{\sigma}(\varepsilon,{\bold R_{m}})],
\end{equation}
where
\begin{equation}
\tilde{\mathcal{G}_{\sigma}}(\varepsilon,{\bold R_{m}})=\frac{1}{\mathcal{N}}\sum_{\mathbf{k}\mathbf{q}}\sum_{\tau\tau'}e^{-i\mathbf{k}\cdot\mathbf{R}_{m}}e^{i\mathbf{q}\cdot\mathbf{R}_{m}}\tilde{\mathcal{G}}_{c_{\mathbf{k}\tau\sigma}c_{\mathbf{q}\tau'\sigma}}\label{eq:7}
\end{equation}
is system's Green function in energy domain $\varepsilon$ at the STM-tip position $\mathbf{R}_{m}.$
By applying the equation-of-motion (EOM) procedure\cite{EOM,BIC2} for the previous equation, we find
\begin{eqnarray}
\tilde{\mathcal{G}}_{c_{\mathbf{k}\tau\sigma}c_{\mathbf{q}\tau'\sigma}} & = & \frac{(\varepsilon\pm\hbar v_{F}\tau k_{z})\delta_{\mathbf{kq}}\delta_{\tau\tau'}}{\varepsilon^{2}-(\tau\hbar v_{F}k)^{2}}\nonumber \\
 & + & \frac{(\varepsilon\pm\hbar v_{F}\tau k_{z})\sum_{j}V_{j\mathbf{k}}}{\varepsilon^{2}-(\tau\hbar v_{F}k)^{2}}\tilde{\mathcal{G}}_{d_{j\sigma}c_{\mathbf{q}\tau'\sigma}}\nonumber \\
 & + & \frac{(\hbar v_{F}\tau k_{-})\sum_{j}V_{j\mathbf{k}}}{\varepsilon^{2}-(\tau\hbar v_{F}k)^{2}}\tilde{\mathcal{G}}_{d_{j\bar{\sigma}}c_{\mathbf{q}\tau'\sigma}},\label{eq:Pr1-1-1}
\end{eqnarray}
where $\pm$ stands for $\sigma=\uparrow,\downarrow$, respectively with $\bar{\sigma}=-\sigma$ and $k_{\pm}=k_{x}\pm ik_{y}.$ To finish the LDOS evaluation, we first
perform the summation over $\tau$ and $\tau'$, which gives:
\begin{equation}
\tilde{\mathcal{G}}_{c_{\mathbf{k}\sigma}c_{\mathbf{q}\sigma}}^{\text{{full}}}=\frac{2\varepsilon\delta_{\mathbf{kq}}}{\varepsilon^{2}-(\hbar v_{F}k)^{2}}+\frac{2\varepsilon\sum_{j}V_{j\mathbf{k}}}{\varepsilon^{2}-(\hbar v_{F}k)^{2}}\sum_{\tau'}\mathcal{\tilde{\mathcal{G}}}_{d_{j\sigma}c_{\mathbf{q}\tau'\sigma}},\label{eq:PRsigma}
\end{equation}
where we defined $\tilde{\mathcal{G}}_{\text{{AB}}}^{\text{{full}}}\equiv\sum_{\tau\tau'}\tilde{\mathcal{G}}_{\text{{AB}}}$.
By applying the EOM method for the mixed Green function $\sum_{\tau'}\mathcal{\tilde{\mathcal{G}}}_{d_{j\sigma}c_{\mathbf{q}\tau'\sigma}},$
we determine
\begin{eqnarray}
\sum_{\tau'}\mathcal{\tilde{\mathcal{G}}}_{d_{j\sigma}c_{\mathbf{q}\tau'\sigma}} & = & \frac{2\varepsilon\sum_{l}V_{l\mathbf{q}}^{*}}{\varepsilon^{2}-(\hbar v_{F}q)^{2}}\tilde{\mathcal{G}}_{d_{j\sigma}d_{l\sigma}}\label{eq:dcsigma}
\end{eqnarray}
and consequently,
\begin{eqnarray}
\mathcal{\tilde{\mathcal{G}}}_{\sigma}(\varepsilon,{\bold R_{m}}) & = & \frac{1}{\mathcal{N}}\sum_{\mathbf{k}\mathbf{q}}\frac{2\varepsilon\delta_{\mathbf{kq}}}{\varepsilon^{2}-(\hbar v_{F}k)^{2}}+\frac{1}{v_{0}^{2}}\sum_{jl}\tilde{\mathcal{G}}_{d_{j\sigma}d_{l\sigma}}\nonumber \\
 & \times & \Sigma({\bold R_{mj}})\Sigma({\bold R_{lm}}),\label{eq:mainGF}
\end{eqnarray}
in which  $\mathbf{R}_{mj}=\mathbf{R}_{m}-\mathbf{R}_{j}$, $\mathbf{R}_{mj}=-\mathbf{R}_{jm}$ and
\begin{eqnarray}
\Sigma({\bold R_{mj}}) & = & \frac{2v_{0}^{2}}{\mathcal{N}}\sum_{\mathbf{k}}\frac{\varepsilon e^{i\mathbf{k}\cdot\mathbf{R}_{mj}}}{\varepsilon^{2}-(\hbar v_{F}k)^{2}}\label{eq:SE-1}
\end{eqnarray}
gives the non-interacting self-energy. After performing the sum over $\mathbf{k}$ and introducing the energy cutoff $D$ as the band-half width of the 3D-DSM, we get:
\begin{eqnarray}
\Sigma({\bold R_{mj}}) & = & -\frac{3\pi v_{0}^{2}\varepsilon}{D^{3}}\frac{\hbar v_{F}}{|R_{mj}|}\exp\left(\frac{i|R_{mj}|\varepsilon}{\hbar v_{F}}\right).\label{eq:SEjl}
\end{eqnarray}
This equation holds in the domain $\frac{|R_{mj}|\varepsilon}{\hbar v_{F}}\gg1$,
i.e, for long-range positions. Particularly at the adatom site, the self-energy reads
\begin{eqnarray}
\Sigma(0) & = & -\frac{6\varepsilon v_{0}^{2}}{D^{2}}\left(1-\frac{\varepsilon}{2D}\ln\left|\frac{D+\varepsilon}{D\text{-}\varepsilon}\right|\right)-i\pi\text{\ensuremath{\mathcal{D}}}_{0}v_{0}^{2},\label{eq:SE}
\end{eqnarray}
with 3D-DSM density of states (DOS) determined by $\text{\ensuremath{\mathcal{D}}}_{0}=\frac{\Omega}{\pi^{2}\hbar^{3}v_{F}^{3}\mathcal{N}}\varepsilon^{2}=\frac{3\varepsilon^{2}}{D^{3}},$
which exhibits quadratic scaling on energy in agreement with Ref.\cite{DOS}.

As a result the LDOS of the system is given by
\begin{eqnarray}
\text{{LDOS}}(\varepsilon,{\bold R_{m}}) & = & 2\text{\ensuremath{\mathcal{D}}}_{0}+\sum_{jl}\Delta\text{{LDOS}}_{jl}({\bold R_{m}}),\label{eq:LDOS}
\end{eqnarray}
where
\begin{align}
\Delta\text{{LDOS}}_{jl}({\bold R_{m}}) & =-\frac{1}{\pi v_{0}^{2}}\sum_{\sigma}{\tt Im}\{\Sigma({\bold R_{mj}})\tilde{\mathcal{G}}_{d_{j\sigma}d_{l\sigma}}\Sigma({\bold R_{lm}})\},\nonumber \\
 \label{eq:DLDOS}
\end{align}
stands for the term induced by the presence of the buried adatoms. Diagonal terms in it with
$j=l$ describe the electronic waves scattered by individual adatoms, while mixing terms with $j\neq l$ correspond to the waves that travel back and forth between two adatoms. The aforementioned quantities are of major importance for the appearance of the so-called BICs, which emerge when $\Delta\text{{LDOS}}_{jl},$ for $j\neq l$, contribute with  Fano antiresonance\cite{Fano1,Fano2} phase shifted by $\pi$ with respect to the resonance arising from $\Delta\text{{LDOS}}_{jj}.$ Noteworthy, both quantities depend on the DOS of the adatoms
\begin{equation}
\text{DOS}_{jl}=-\frac{1}{\pi}{\tt Im}(\sum_{\sigma}\tilde{\mathcal{G}}_{d_{j\sigma}d_{l\sigma}}).
\end{equation}

To evaluate functions $\tilde{\mathcal{G}}_{d_{j\sigma}d_{l\sigma}}$, we start employing the EOM approach which gives:
\begin{eqnarray}
(\varepsilon-\varepsilon_{d_{j\sigma}}-\Sigma(0))\tilde{\mathcal{G}}_{d_{j\sigma}d_{l\sigma}} & = & \delta_{jl}+U_{j}\tilde{\mathcal{G}}_{d_{j\sigma}n_{d_{j\bar{\sigma}}}d_{l\sigma}}\nonumber \\
 & + & \Sigma({\bold R_{j\bar{j}}})\mathcal{\tilde{\mathcal{G}}}_{d_{\bar{j}\sigma}d_{l\sigma}},\label{eq:Gdd1}
\end{eqnarray}
where $\bar{j}=1,2$ when $j=2,1$. In this expression, $\tilde{\mathcal{G}}_{d_{j\sigma}n_{d_{j\bar{\sigma}}}d_{l\sigma}}$
stands for two-particle Green function, which yields
\begin{align}
(\varepsilon-\varepsilon_{d_{j\sigma}}-U_{j})\mathcal{\tilde{\mathcal{G}}}_{d_{j\sigma}n_{d_{j\bar{\sigma}}}d_{l\sigma}} & =\delta_{jl}\left\langle n_{d_{j\bar{\sigma}}}\right\rangle \nonumber \\
+\sum_{\mathbf{k}\tau}V_{j\mathbf{k}}[\tilde{\mathcal{G}}_{d_{j\bar{\sigma}}^{\dagger}c_{\mathbf{k}\tau\bar{\sigma}}d_{j\sigma},d_{l\sigma}} & +\tilde{\mathcal{G}}_{c_{\mathbf{k}\tau\sigma}d_{j\bar{\sigma}}^{\dagger}d_{j\bar{\sigma}},d_{l\sigma}}\nonumber \\
-V_{j\mathbf{k}}\tilde{\mathcal{G}}_{c_{\mathbf{k}\tau\bar{\sigma}}^{\dagger}d_{j\bar{\sigma}}d_{j\sigma},d_{l\sigma}} & ],\label{eq:Gdnd-1}
\end{align}
where the occupation number is
\begin{eqnarray}
\left\langle n_{d_{j\bar{\sigma}}}\right\rangle  & = & -\frac{1}{\pi}\int_{-D}^{0}{\tt Im}(\tilde{\mathcal{G}}_{d_{j\bar{\sigma}}d_{j\bar{\sigma}}})d\varepsilon.
\end{eqnarray}
We employ the Hubbard I approximation\cite{HubbardI,BIC2} in order to close this
dynamic set of the equations for Green functions. Thereby, we find for the diagonal adatom Green functions
\begin{eqnarray}
\tilde{\mathcal{G}}_{d_{j\sigma}d_{j\sigma}} & = & \frac{\lambda_{j}^{\bar{\sigma}}}{\varepsilon-\varepsilon_{d_{j\sigma}}-\tilde{\Sigma}_{jj}^{\sigma}},\label{eq:Gdd}
\end{eqnarray}
where $\lambda_{j}^{\bar{\sigma}}=1+\frac{\left\langle \right.n_{d_{j\bar{\sigma}}}\left.\right\rangle U_{j}}{\varepsilon-\varepsilon_{d_{j\sigma}}-U_{j}-\Sigma(0)}$
and {$\tilde{\Sigma}_{jj}^{\sigma}=\Sigma(0)+\lambda_{j}^{\bar{\sigma}}\lambda_{\bar{j}}^{\bar{\sigma}}\frac{\Sigma({\bold R_{j\bar{j}}})\Sigma({\bold R_{\bar{j}j}})}{\varepsilon-\varepsilon_{d_{\bar{j}\sigma}}-\Sigma(0)}=\Sigma(0)+\Sigma_{jj}^{\sigma}.$}
The mixed Green functions are:

\begin{eqnarray}
\tilde{\mathcal{G}}_{d_{j\sigma}d_{\bar{j}\sigma}} & = & \frac{\lambda_{j}^{\bar{\sigma}}\Sigma({\bold R_{j\bar{j}}})}{\varepsilon-\varepsilon_{d_{j\sigma}}-\Sigma(0)}\mathcal{\tilde{G}}_{d_{\bar{j}\sigma}d_{\bar{j}\sigma}}.\label{eq:Gddjl-1-1}
\end{eqnarray}

\begin{figure}[!]
\centering{}\includegraphics[scale=0.32]{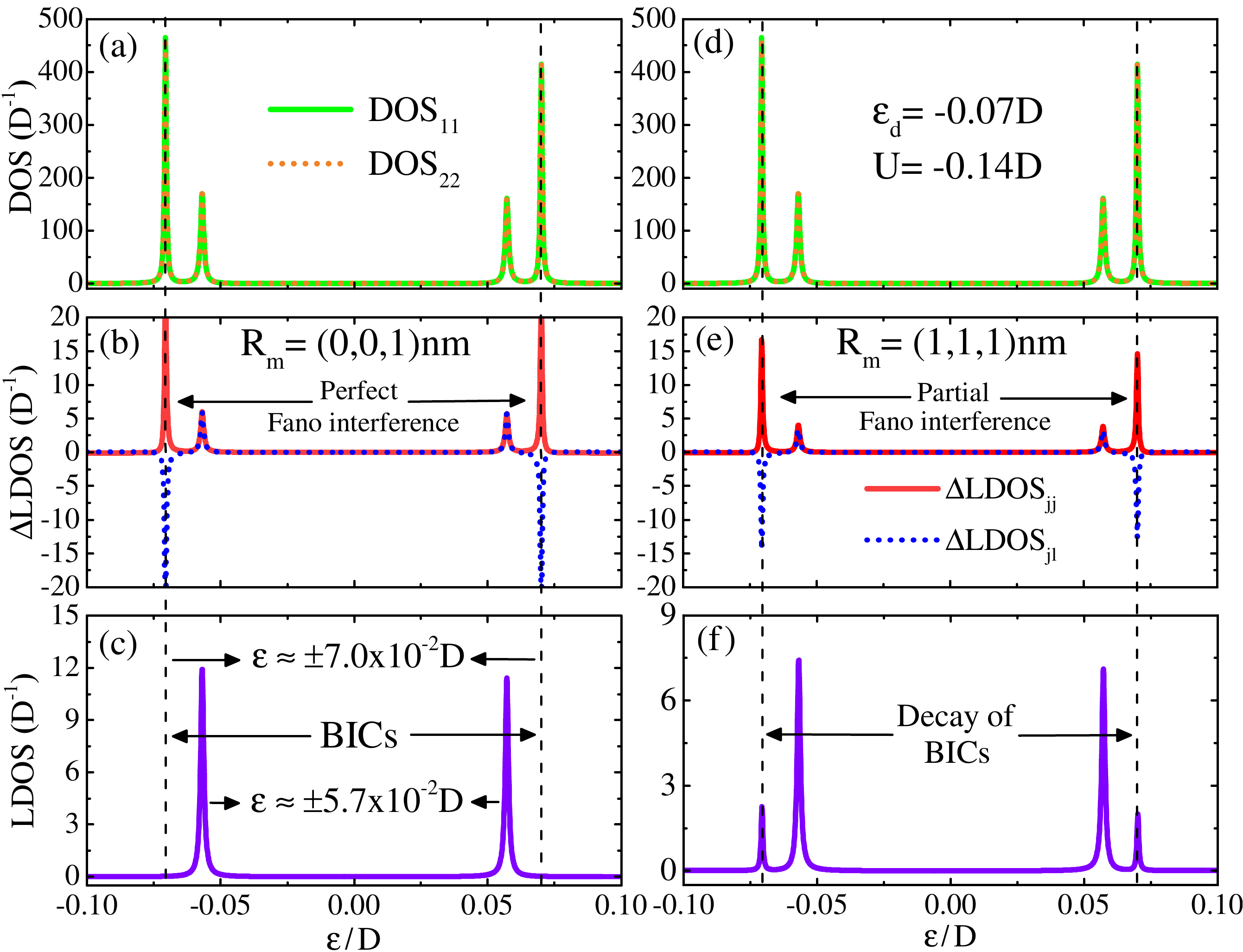}\protect\protect\protect\caption{\label{fig:Pic2}(Color online) (a) and (d) $\text{DOS}_{11}$ (solid
green curve) and $\text{DOS}_{22}$ (dashed orange curve) of the adatoms. (b) $\Delta\text{LDOS}_{jl}$
profiles at $\mathbf{R}_{m}=(0,0,1)\text{{nm}}$, described by Eq.(\ref{eq:DLDOS}), for $j=l$ (solid red curve) and $j\protect\neq l$
(dashed blue curve), where the resonances are perfectly canceled
by Fano antiresonances around $\varepsilon\approx\pm7.0\times10^{-2}D$. (c)
Total LDOS revealing BICs on the surface of the 3D-DSM host. (e) $\Delta\text{LDOS}_{jl}$
profiles at $\mathbf{R}_{m}=(1,1,1)\text{{nm}}$ where the destructive
interference is not perfect. (f) Total LDOS revealing the decay of BICs at the same energetic positions where BICs are found. The
vertical dashed lines crossing panels indicate the positions where
the Fano destructive interference processes occur.}
\end{figure}

\textit{Results and Discussion.}---
In the following discussion we consider the case of  two identical adatoms placed at $\mathbf{R}_{1,2}=(0,\mp1,0)\text{{nm}}$ (the surface of the system corresponds to $(x,y,1)nm$ plane),
with energy levels $\varepsilon_{d_{j\sigma}}=-0.07D$, which are
hybridized to the free electrons of 3D-DSM with strength $v_{0}=0.14D$ and on-site Coulomb repulsion $U_{j}=0.14D$. {We point out that the change of $v_{0}$ just shifts rigidly the profile of the adatom DOS.} Additionally, we have
chosen $\hbar v_{F}\approx5\:eV\text{\AA}$ and $D\approx0.2\,\text{\text{{eV}}}$,
which are experimental parameters for Cadmium Arsenide ($\text{Cd}_{3}\text{As}_{2}$)\cite{DSM3,DSM5}. The set of parameters we use corresponds to symmetric Anderson regime ($2\varepsilon_{d_{j\sigma}}+U_{j}=0$). For such conditions the Hamiltonian becomes invariant under particle-hole transformation as can be seen in Fig.\ref{fig:Pic2}. The presence of the particle-hole symmetry is no way necessary for the appearance of the phenomena discussed below.

The four-peak structure in DOS visible in the upper panels of Fig.\ref{fig:Pic2} emerges from the contributions provided by Coulomb repulsion $U_{j}$ and interacting self-energy {${\Sigma}_{jj}^{\sigma}=\lambda_{j}^{\bar{\sigma}}\lambda_{\bar{j}}^{\bar{\sigma}}\frac{\Sigma({\bold R_{j\bar{j}}})\Sigma({\bold R_{\bar{j}j}})}{\varepsilon-\varepsilon_{d_{\bar{j}\sigma}}-\Sigma(0)}$} in Eq.(\ref{eq:Gdd}) for the adatoms Green functions. The former leads to the formation of the pair of peaks at $\varepsilon_{d_{j\sigma}}$ and $\varepsilon_{d_{j\sigma}}+U_{j}$ as expected, the latter is responsible for the splitting of both of them. Noteworthy, this self-energy provides effective tunneling between the adatoms mediated by the bulk states of the 3D-DSM, even in the absence of the direct tunneling term $t({\bold R_{12}})d_{1\sigma}^{\dagger}d_{2\sigma}+\text{H.c.}$. This indirect tunneling becomes specially important when adatoms are well separated from each other. This four-peak structure corresponds to the formation of molecular states with remarkable property: the ground-state corresponds to the antibonding configuration. This is consequence of the particular scaling of 3D-DSM DOS with energy $\text{\ensuremath{\mathcal{D}}}_{0}\propto\varepsilon^{2}$ entering into expression for $\Sigma({\bold R_{mj}})$ as a result. If we replace this DOS by the one corresponding to the normal metal the reported effect disappears. {Additionally, following Ref.\cite{SOC} and looking at the poles of the adatom Green function, we recognize $t_{\text{eff}}=\text{Re}(\Sigma(0)+{\Sigma}_{jj}^{\sigma})$ as the effective hopping term between the adatoms, which is negative as we have checked it, thus ensuring the antibonding ground state. However, distinctly from Ref.\cite{SOC} where the negative tunneling term comes from the spin-orbit coupling, here it emerges from Friedel-like oscillations inside the relativistic 3D-DSM environment encoded by the self-energy ${\Sigma}_{jj}^{\sigma}$.}

The nature of the four molecular states can be clarified if one analysis the corresponding LDOS of the whole system. Note that, this quantity is position dependent and its profile on the surface of the system can be visualized experimentally using an STM-tip.   Middle panels at Fig.\ref{fig:Pic2} illustrate the contribution of the adatoms on the surface LDOS evaluated at $\mathbf{R}_{m}=(0,0,1)\text{\text{{nm}}}$ (Fig.\ref{fig:Pic2}(b)) and $\mathbf{R}_{m}=(1,1,1)\text{\text{{nm}}}$ (Fig.\ref{fig:Pic2}(e)). In both panels diagonal terms ($j=l$) present pronounced peaks at the same energies as those of the DOS (upper panels of Fig.\ref{fig:Pic2}). The mixed terms ($j\neq l$) show resonances around $\varepsilon\approx\pm5.7\times10^{-2}D$ and antiresonances nearby $\varepsilon\approx\pm7.0\times10^{-2}D$. When one computes the total LDOS as sum of all contributions from $\Delta\text{{LDOS}}_{jl}$ the interference between diagonal and mixed terms can be constructive or destructive. For the latter case the peaks in total LDOS become attenuated and can even totally vanish as it happens at Fig.\ref{fig:Pic2}(c), where only two peaks out of four survive. In this case two peaks disappearing due to Fano destructive interferences\cite{Fano1,Fano2} around $\varepsilon\approx\pm7.0\times10^{-2}D$, correspond to the BICs. Note that, full annihilation of the peaks takes place only for certain values of $\mathbf{R}_{m}$ as one can clearly see at the Figs.\ref{fig:Pic2}(e) and (f). In this case the destructive interference is not perfect and thus BICs inevitably experience decay into the host continuum.

\begin{figure}[!]
\centering{}\includegraphics[scale=0.43]{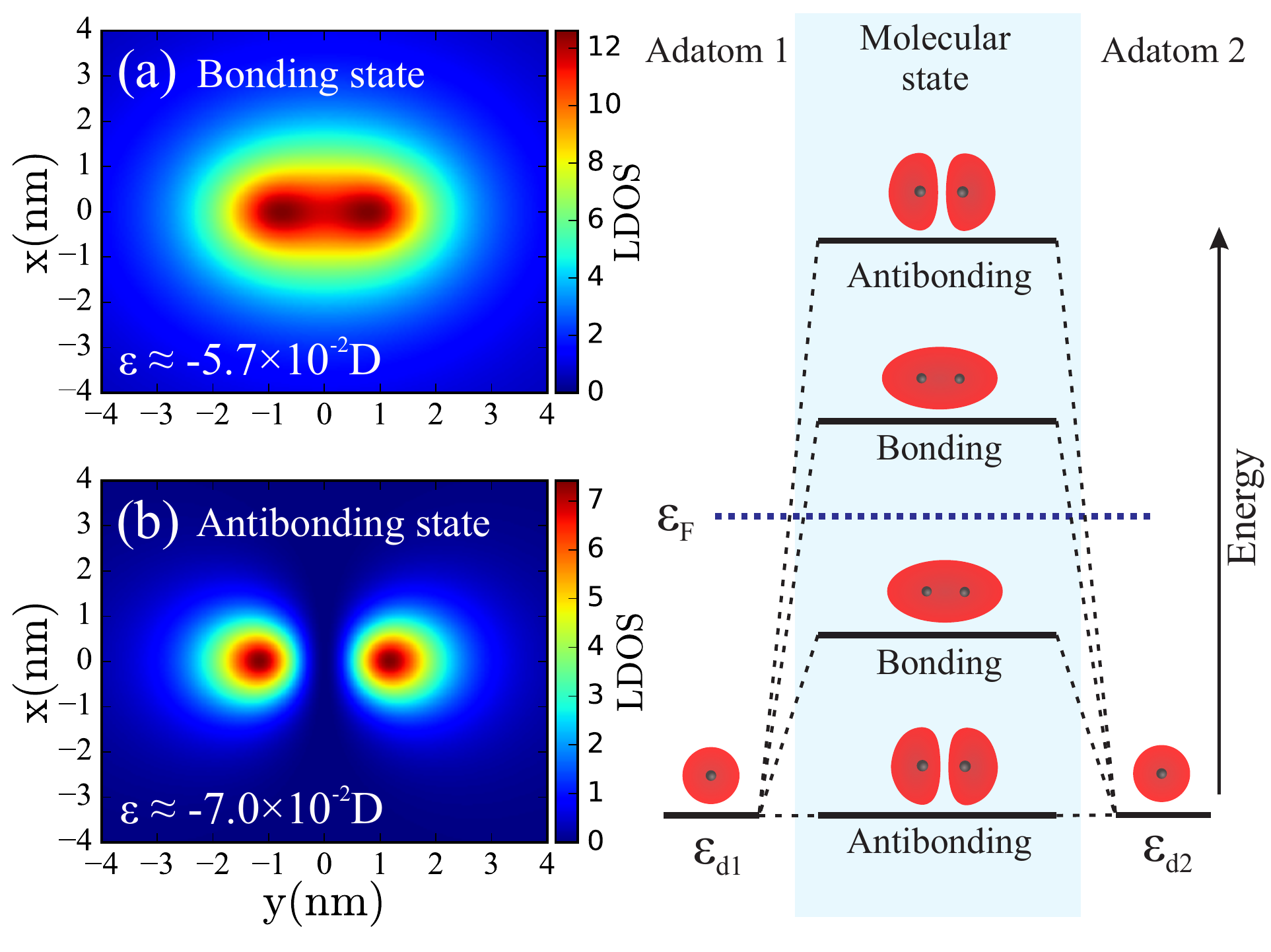}
\protect\protect\protect\caption{\label{fig:Pic3}(Color online) Topography of $\Delta\text{LDOS}$
on the surface of 3D-DSM ($\mathbf{R}_{m}=(x,y,1)\text{\text{{nm}}}$ plane) for two relevant
energy values: (a) $\varepsilon\approx-5.7\times10^{-2}D$ corresponding to the constructive interference of the diagonal and mixed terms in $\Delta\text{{LDOS}}$. One can clearly see the bonding character of the density profile. (b) $\varepsilon\approx-7.0\times10^{-2}D$ corresponding to the destructive interference of the diagonal and mixed terms in $\Delta\text{{LDOS}}$ for which an antibonding molecular state emerges. Note that, this state corresponds to the ground state of the system (c) A scheme of the hierarchy of the molecular states.}
\end{figure}

\begin{figure}
\centering{}\includegraphics[scale=0.43]{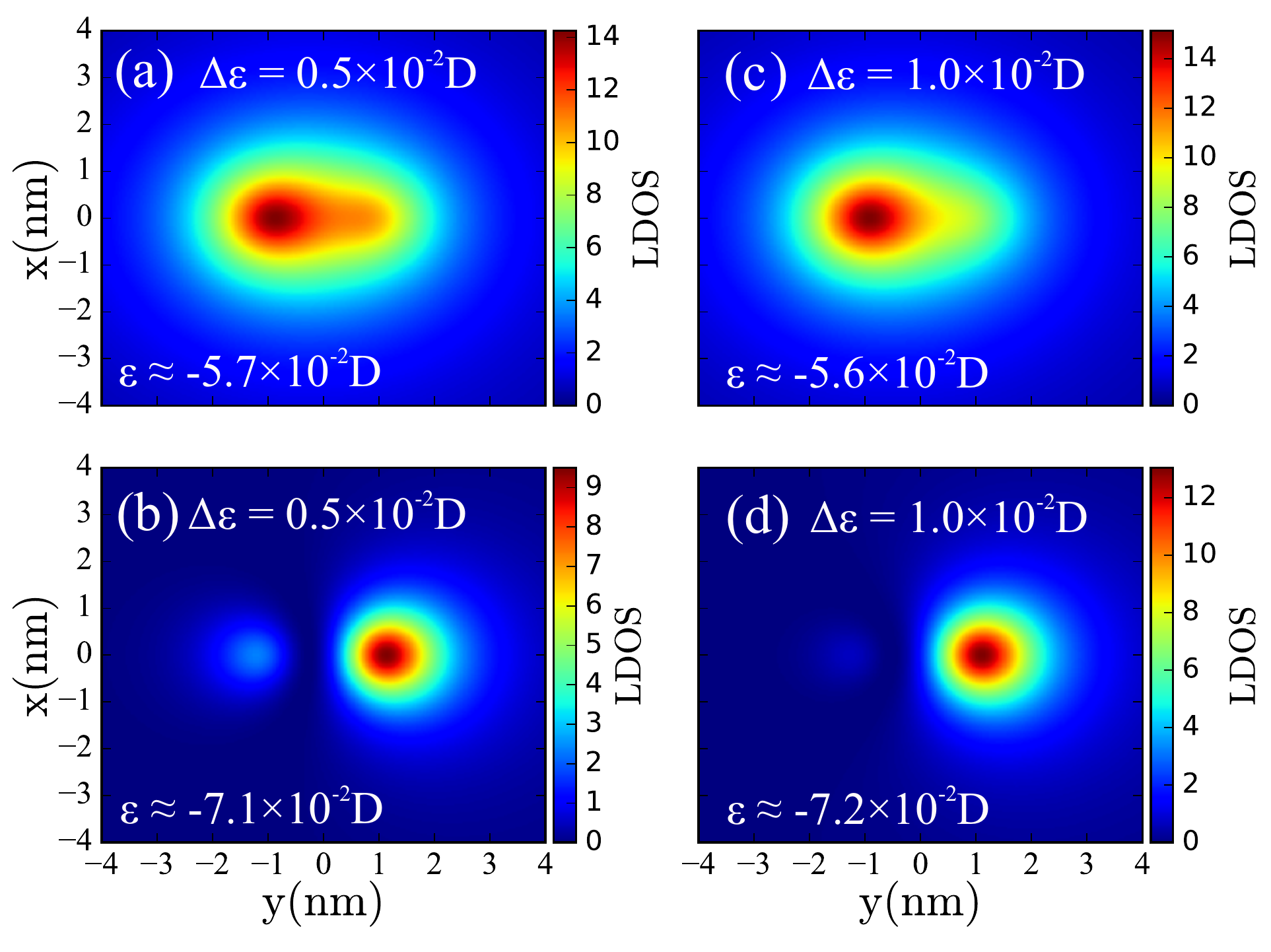}
\protect\protect\protect\caption{\label{fig:Pic4}(Color online) Topography of $\Delta\text{LDOS}$
on the surface of 3D-DSM ($\mathbf{R}_{m}=(x,y,1)\text{\text{{nm}}}$ plane) illustrating the effects of energy detuning between the adatom levels. Panels (a) and (b) show bonding and antibonding profiles for $\Delta\varepsilon=0.5\times10^{-2}D$ corresponding to the energies $\varepsilon\approx-5.7\times10^{-2}D$ and $\varepsilon\approx-7.0\times10^{-2}D$, respectively. Panels (c) and (d) show bonding and antibonding profiles for $\Delta\varepsilon=1.0\times10^{-2}D$ corresponding to the energies $\varepsilon\approx-5.6\times10^{-2}D$ and $\varepsilon\approx-7.2\times10^{-2}D$ respectively.}
\end{figure}

The profiles of the total LDOS on 3D-DSM surface ($\mathbf{R}_{m}=(x,y,1)\text{\text{{nm}}}$ plane) are shown in Fig.\ref{fig:Pic3}. We considered  two distinct values of the energies corresponding to the cases of constructive and destructive Fano interference in $\Delta\text{{LDOS}}$, (i) $\varepsilon\approx-5.7\times10^{-2}D$,
(ii) $\varepsilon\approx-7.0\times10^{-2}D$ respectively. For the case of constructive interference shown at Fig.\ref{fig:Pic3}(a) density profile reveals nodeless covalent molecular state, i.e., bonding state. On the contrary, when the energy corresponds to destructive Fano interference and formation of BIC, the density profile has pronounced node between the adatoms and thus corresponds to the antibonding state. Note that, this latter case corresponds to the peak in DOS with minimal energy. Thus, differently from the case of the real molecules the ground molecular state is antibonding, which is quite remarkable\cite{Experiment}. It is worth mentioning that another pair of  bonding and antibonding states exist above the Fermi energy ($\varepsilon_{F}=0$) due to the particle-hole symmetry of the original Hamiltonian (Fig.\ref{fig:Pic3}(c)).

The molecular states discussed above are robust with respect to the detuning $\Delta\varepsilon$ of the energy levels of two adatoms. The corresponding profiles of LDOS for bonding and antibonding states are shown in the Fig.\ref{fig:Pic4} for two different values of $\Delta\varepsilon$. Naturally, profiles become asymmetric, but nodal line between the adatoms revealing the antibonding  nature of the ground-state remains clearly visible.

\textit{Conclusions.}---To summarize, we evaluated the LDOS on the surface of 3D-DSM hosting two distant buried adatoms and found that the ground state of this molecular system has density profile with node between the atoms and thus corresponds to the antibonding state. This is in contrast with natural molecules for which the ground state is always bonding. The predicted effect appears due to the indirect tunneling between the adatoms mediated by quasi-relativistic free electrons of 3D-DSM.

\textit{Acknowledgments.}---This work was supported by the agencies CNPq (307573/2015-0), CAPES, S{ã}o Paulo Research Foundation (FAPESP) - grant: 2015/23539-8. I.A.S. acknowledges the support from Horizon2020 RISE project CoExAN and RSF(17-12-01581). A.E.O. thanks to CNPq (grant: 312838/2016-6) and Secti/FAPEMA (DCR 02853/16).

\end{document}